\documentclass[final,1p,times]{elsarticle}

\usepackage{amssymb}

\usepackage{amsmath}
\usepackage{graphicx}
\usepackage{natbib}
\usepackage{color}
\usepackage{multirow}
\definecolor{Dark}{gray}{.90}

\journal{Neuroscience Letters}

\begin{document}

\begin{frontmatter}



\title{Point-process deconvolution of fMRI reveals effective connectivity alterations  in chronic pain patients }


\author{Guo-Rong Wu$^{1}$, Enzo Tagliazucchi$^{3}$, Dante R. Chialvo$^{4,5}$ Daniele Marinazzo$^{1}$}

\address{ $^{1}$ Faculty of Psychology and Educational Sciences, Department of Data Analysis,
Ghent University, Henri Dunantlaan 1, B-9000 Ghent, Belgium   \\
$^{1}$ Key Laboratory for NeuroInformation of Ministry of Education, School of Life Science and Technology, University of Electronic Science and Technology of China, Chengdu, China \\
$^{3}$ Neurology Department and Brain Imaging Center, Goethe University, Frankfurt am Main, Germany \\
$^{4}$ Consejo Nacional de Investigaciones Cient\'ificas y Tecnol\'ogicas (CONICET), Buenos Aires, Argentina \\
$^{5}$ Facultad de Ciencias M\'edicas, Universidad Nacional de Rosario, Rosario 2000, Argentina \\
}

\begin{abstract}
It is now recognized that important information can be extracted from the brain spontaneous activity, as exposed by recent analysis using  a repertoire of computational methods. In this context a novel  method, based on a blind deconvolution technique, is used in this paper to analyze potential changes due to chronic pain in the brain pain matrix's effective connectivity. 
The approach is able to deconvolve the hemodynamic response function to spontaneous neural events, i.e., in the absence of explicit onset timings, and to evaluate information transfer between two regions as a joint probability of the occurrence of such spontaneous events.
The method revealed that the chronic pain patients exhibit important changes in the Insula's effective connectivity which can be relevant to understand the overall impact of chronic pain on brain function.

\end{abstract}


\end{frontmatter}


\section{Introduction}

Recent results shows that chronic pain is  a condition that, beyond the feeling of acute pain, affects normal brain function and structure, causing cognitive impairments, including depression, sleeping disturbances and decision-making abnormalities \cite{apka2004,apka2005,baliki2008}. 

Disturbances in cortical dynamics due to chronic pain have been demonstrated using functional magnetic resonance imaging (fMRI), both studying activation in response to  external stimulation\cite{derby1999,peyron2000} as well as using seed based correlation analysis during the execution of simple attention demanding tasks \cite{baliki2008}. In particular, the later study showed for the first time that the dynamics of the default mode network (DMN) is disrupted in chronic pain.

More recent studies \cite{baliki2012} show that it is even possible to identify a temporal profile of brain parameters which changes during pain chronification in patients suffering sub acute back pain.  These changes involve media prefrontal regions of the cortex, the Insula as well as the Nucleous Acumbens. The insular cortex is often activated bilaterally  during noxious somatosensory stimulation and has been suggested to play an important role in pain processing \cite{coghill1994,coghill1999}. At the same time, the extensive connectivity of the Insula suggests a multifaceted role in the dynamic of pain perception, and the need to develop new methods to unravel its complexity. 

The present study uses a novel approach to detect neural events in BOLD signals to investigate the network of {\it directed} dynamical influences between brain regions involved in pain processing, in particular the Insular region. The approach is able to deconvolve the hemodynamic response function (HRF) to spontaneous neural events, i.e., in the absence of explicit onset timings, and to evaluate information transfer between any two regions as a joint probability of the occurrence of such spontaneous events.

The paper is organized as follows: The next section describe the brain imaging data as well as the numerical methods for deconvolution and Granger causality mapping.  In Section 3 the main findings are presented  indicating important changes in the Insula's effective connectivity in chronic pain patients. The paper closes in Section 4 with a brief discussion on the method novelty as well as on the physiological relevance of the results. 

\section{Materials and methods}

\subsection{fMRI data acquisition and preprocessing}
The data analyzed here corresponds to 12 chronic back pain patients (CBP) (range 29-67 years old, mean=51.2) and 20 healthy controls (HC) (range 21-60 years old, mean=38.4). All subjects were right-handed and all gave informed consent to procedures approved by Northwestern University (Chicago) IRB committee \cite{tagliazucchi2010a}.

Participants were asked to lay still in the scanner and to keep their mind blank, eyes closed and avoid falling asleep \cite{fox2005}. Functional magnetic resonance data  was acquired using a 3T Siemens Trio whole-body scanner with echo-planar imaging capability using the standard radio-frequency head coil. Scanner parameters were similar to those used in an earlier study\cite{baliki2008}. For each subject, a total of 300 images (spaced by 2.5 sec) were obtained, in which the brain oxygen level dependent (BOLD) signal was recorded for each one of the 64 $\times$ 64 $ \times $ 49 sites (voxels of dimension 3.4375 mm $\times$ 3.4375 mm $\times$ 3 mm).
 
Preprocessing of BOLD signal was performed using FMRIB Expert Analysis Tool (\cite{jezzard} ; http://www.fmrib.ox.ac.uk/fsl). Data preprocessing included motion correction using MCFLIRT, slice-timing correction using Fourier-space time-series phase-shifting, non-brain removal using BET, spatial smoothing using a Gaussian kernel of full-width-half-maximum 5 mm. Brain images were normalized to standard space using the MNI 152  template using FLIRT and data was resampled to 4 mm $\times$ 4 mm $\times$ 4 mm resolution. A zero lag finite impulse response filter was applied to band pass filter  (0.01 Hz - 0.1 Hz) the functional data (the lower frequency was chosen to avoid noise related to scanner drift and the higher frequency was chosen to eliminate  high frequency artifacts related with physiological noise and head motion) \cite{cordes2000,cordes2001}. An independent component analysis  (ICA) de-noising procedure   \cite{beckmann2004} consisting of edge removal and high frequency artifacts by linear regression was performed using Melodic. 
	
A predefined pain matrix mask was employed in the present study, already described in previous works \cite{tagliazucchi2010a,tagliazucchi2010b}. As a control, a region with no expected pain effects, the primary visual cortex (BA 17) was used  for comparison.

\subsection{Spontaneous point event detection and HRF Deconvolution}

\begin{figure}[h]
\begin{center}
\includegraphics[scale=3]{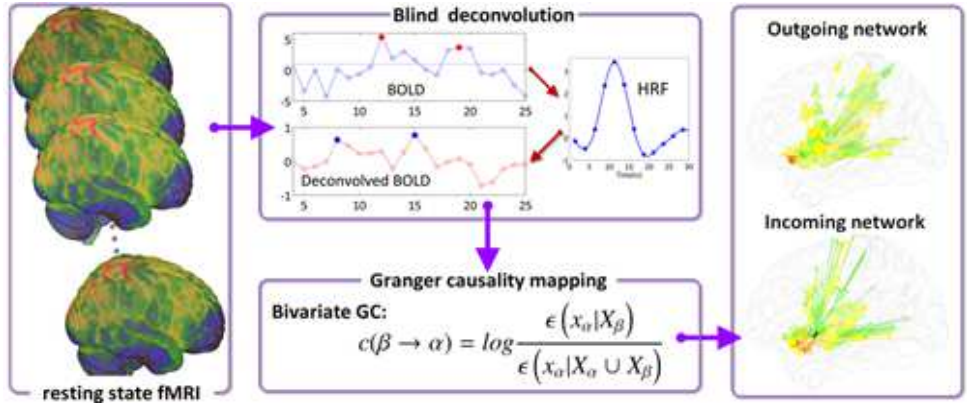}
\caption{Flow chart of the procedure: a blind deconvolution algorithm is applied to BOLD signals. Granger causality analysis is performed on the deconvolved BOLD signals to obtain the network of incoming and outgoing influences. Furthermore for each voxel the HRF is retrieved.}
\end{center}
\label{figmethods}
\end{figure}

Previous studies have shown that the hemodynamic processes are inhomogeneous across the whole brain \cite{handwerker2012}. These inhomogeneities acting  over the hemodynamic response can limit the inferences of temporal precedence \cite{valdes-sosa2011} which are central for establishing effective connectivity between regions. To overcome this limitation, a novel blind deconvolution technique (see Fig.\ref{figmethods} was developed recently for resting-state BOLD-fMRI signals \cite{wu2013}. The approach relies on the idea that the resting-state BOLD spikes can be seen as the response to  spontaneous  neuronal events, something supported by the increasing evidence of non-random patterns governing the dynamics of the brain at rest \cite{tagliazucchi2011,petridou2012}. 

These spontaneous events can be detected by point process analysis (PPA), picking up BOLD fluctuations of relatively large amplitude \cite{tagliazucchi2010a,tagliazucchi2010b,tagliazucchi2012}. After detecting these resting-state BOLD transients, the BOLD event onsets are stored for further reconstruction of the hemodynamic response function. The voxel-specific' HRF is obtained by fitting raw BOLD signal with triggered averages and shifted BOLD event onsets, in order to finally recover signals at the neural level by Wiener deconvolution \cite{glover1999}. 

To characterize the hemodynamic response function (HRF) elicited by spontaneous point events, two easily interpretable parameters of the HRF which estimate the potential changes in neuronal activity \cite{lindquist2007} were calculated: the response height and the time to peak.

\begin{figure}[htb]
\centering{\includegraphics[scale=1.2]{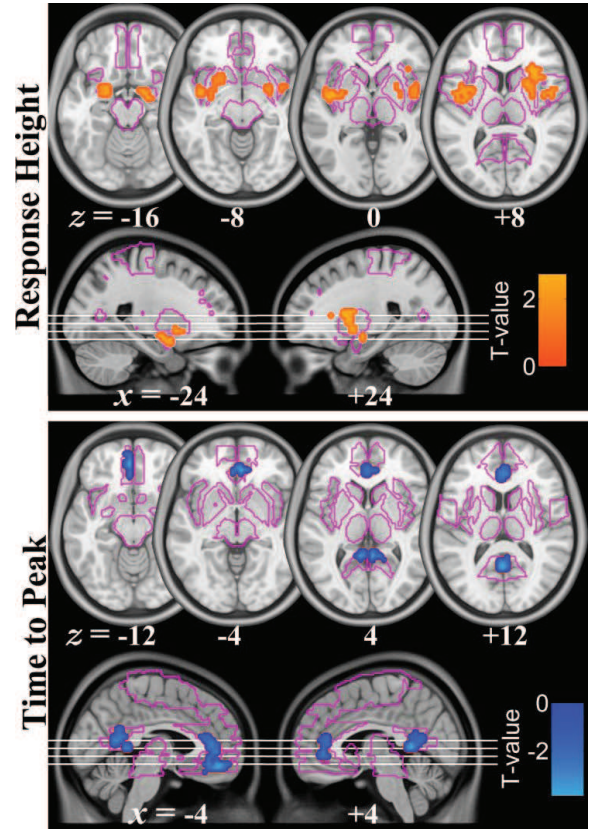}
\caption{Significant differences in the HRF parameters between chronic pain patients versus the control group. Top panels represent the response height and the bottom panels the latency to the peak of the response. Voxels depicted overlaid on the anatomical image are those which passed Alphasim correction with $p < 0.05$. The purple contour lines indicate the location of the voxels belonging to the pain matrix.  
\label{fighrf}}}
\end{figure}
\subsection{Granger Causality mapping}
Given $k$  covariance-stationary variables $\left\lbrace x_i(t)\right\rbrace _{i=1,\cdots,k}$, the state vectors are denoted $X_{\alpha}(t)=\left( x_{\alpha}(t-m),\cdots , x_{\alpha}(t-1) \right)$,
$m$ being the model order. Let $\epsilon(x_{\alpha}|Y)$ be the mean squared error prediction of $x_{\alpha}$ on the basis of the vectors $Y$. The Granger causality index from $\beta\in\Re^{N\times 1}$ to $\alpha\in\Re^{N\times 1}$  is defined as follows:
\begin{equation}
c(\beta \rightarrow \alpha)=
log\frac{\epsilon \left( x_{\alpha} | X_{\beta}\right)}{\epsilon \left( x_{\alpha} | X_{\alpha} \cup X_{\beta}\right)}
\end{equation}
in addition, for further statistical analysis, GC value $c$ is transformed into $c' = \sqrt{n\cdot c - (m-1)/3}$, which is considered to be approximately normal (where $n = N-m$. If $c=0$, $n\cdot c\sim \chi^2(m)$) \cite{geweke1982}.\\

The model order used in this study is $m=1$, evaluated by leave-one-out cross-validation, and common in fMRI GC studies \cite{roebroeck2005}. Regression models were estimated by the ordinary least squares algorithm.\\

Pairwise causal interaction was investigated by mapping the influence between the bilateral Insula and the BOLD time series of the individual voxels belonging to the entire pain matrix (see above).

\begin{figure}[h]
\begin{center}
\includegraphics[scale=1.5]{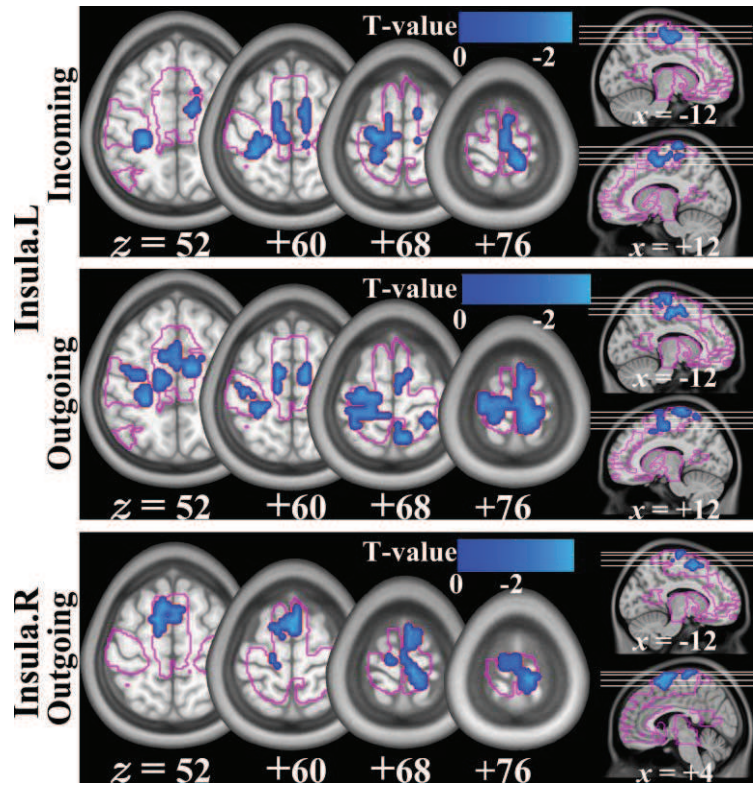}
\caption{Granger Causality mapping between the Insula and the entire pain matrix. Results illustrate the significant differences  between the Granger Causality maps of the patients versus the control group. Voxels depicted, overlaid on the anatomical images, are those passing  Alphasim correction with $p  < 0.05$. As in Fig. \ref{fighrf}, the pain matrix is indicated by the purple contour lines. 
\label{figGC}}
\end{center}
\end{figure}

\subsection{Statistical Analysis}
To compute the group differences (i.e., patient vs. control groups) a two-sample T-test was implemented in SPM, independently for HRF parameters and seed based GC mapping. Statistical significance was estimated via a Monte Carlo simulation (Alphasim, http://afni.nih.gov/afni/docpdf/AlphaSim.pdf). A cluster-wise threshold of $p < 0.05$ by combining a $p < 0.04$ individual voxel threshold and different minimum cluster size of $k$ contiguous voxels (implemented in the REST toolbox, www.restfmri.net; Gaussian filter width was estimated from each SPM T-map, cluster connection radius: 5 mm and 1000 iterations).

\section{Results}
 
\subsection{Spontaneous hemodynamic response}

Figure 1 summarizes the main findings concerning the parameters of spontaneous BOLD activity. It is seen that in terms of the HRF's time to peak (compared to the control subjects) patients response function was characterized by {\it longer time to peak latency} in Precentral and Postcentral Gyrus, but {\it shorter time to peak latency} in Anterior Cingulate, Posterior Cingulate, Medial Frontal Gyrus, Dorsal anterior Cingulate Cortex, Orbitofrontal Cortex, Precuneus and Retrosplenial Cingulate Cortex (see bottom panel of Fig. \ref{fighrf}). 

Concerning the other parameter, the HRF response height, the major modifications were found predominately in the Insula. Other regions included Putamen, Superior Temporal Gyrus, Parahippocampal Gyrus, Caudate and Amygdala (see top panel of Fig.  \ref{fighrf}). As a control we computed similar quantities in the primary visual cortex V1, a region not involved in pain processing, exhibiting no significant difference between groups.

\subsection{Seed-based Granger causality mapping}
To compute the Granger causality mapping we selected seeds ROIs based on the two sample t-test results of the hemodynamic responses. They were centered at the two peaks T-values inside the bilateral insular regions (MNI coordinates: left Insula, [-40 -4 -4]; right Insula, [32 20 8]; with sphere 8mm diameter). GC mapping was then independently implemented for left and right Insula.

The GC mapping between  left Insula and the following regions exhibited lower information transmission (included incoming and outgoing information): Medial Frontal Gyrus, Precentral Gyrus, Postcentral Gyrus, Premotor Cortex, Supplementary Motor Area, Paracentral Lobule, Primary Motor Cortex, Primary Somatosensory Cortex.

For the case of the right Insula it was found that  there is  significantly less information transfer for voxels  located in: Premotor Cortex, Superior Frontal Gyrus, Supplementary Motor Area, Medial Frontal Gyrus, Paracentral Lobule, Precentral Gyrus, Postcentral Gyrus.  No significant difference in the effective connectivity was reported from the pain matrix to right Insula.

The results are presented in figure \ref{figGC} and summarized in table \ref{table1}.

\begin{table}[h]
  \centering
    \begin{tabular}{|c|c|c|c|c|c|c|c|}
    \hline
    Parameter     & Brain region & BA    & C. size & \multicolumn{3}{|c|}{peak MNI~(x,y,z)} & peak T value \\
    \hline
    \multirow{2}[4]{*}{Resp. Height} & Temporal\ Sup\ L & -     & 116   & -44   & -4    & -8    & 3.42 \\
          & Hippocampus\ R & -     & 124   & 20    & -8    & -16   & 3.01 \\
    \hline
    \multirow{3}[6]{*}{Time to Peak} & Postcentral\ L & -     & 62    & -32   & -28   & 72    & 3.97 \\
          & Rectus\ L & 11    & 78    & -4    & 44    & -16   & -4.09 \\
          & Vermis\ 4\ 5 & -     & 52    & 4     & -48   & 4     & -4.64 \\
    \hline
    In, L.I. & Supp\ Motor\ Area\ R & -     & 215   & 12    & -8    & 56    & -2.92 \\
    \hline
    Out, L.I. & Paracentral\ Lobule\ R & -     & 445   & 8     & -28   & 80    & -3.57 \\
    \hline
    Out, R.I. & Paracentral\ Lobule\ R & -     & 166   & 8     & -28   & 80    & -3.93 \\
    \hline
    \end{tabular}
    \\
\caption{Significant results of HRF parameters/Granger Causality mapping resulting from the comparison between the patient and the control groups.
Abbreviations: C. size: Cluster size; Res. Height: Response Height; Time to Peak : Time to Peak response; In, L.I.: Incoming network, Left Insula; Out, L.I.: Outgoing network, Left Insula; Out, R:I. Outgoing network, Right Insula}
  \label{table1}
\end{table}

\subsection{Joint probabilities of neural events}
To further characterize the causal effects from and to the Insula, and to demonstrate how instantaneous neural events detected in the BOLD signal can be not only helpful for deconvolution, but can even reveal themselves this causal effects, we investigated the relative timing of the onset of the neural events in the Paracentral Right Lobule (see table \ref{table1}) with respect to those occurring in the Left Insula. 
 
\begin{figure}[h]
\begin{center}
\includegraphics[scale=2]{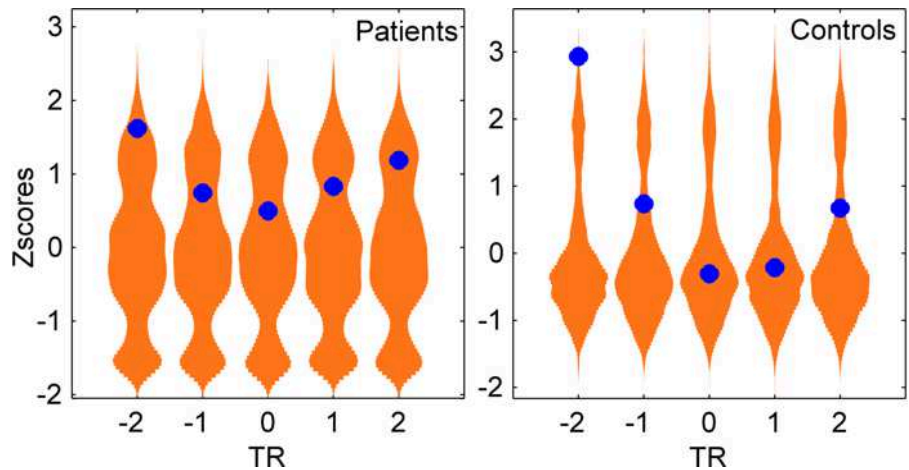}
\caption{Distribution of the relative timing of events occurring  in the Paracentral Right Lobule with respect to events in the Left Insula. The blue dots represent the relative number of large BOLD activations occurring in voxels belonging to the Paracentral Right Lobule each time that a large event is detected in the Left Insula, against the distribution of randomized events for each lag (in orange), for patients (left) and control subjects (right).   \label{fig_prob_onset}  
}
\end{center}
\end{figure}

Every time that an event was detected in the time series of a voxel belonging to the Paracentral Right Lobule, we searched for other events occurring in the Left Insula $[-L,L]$ centered at the onset, where $L = 3 ~TR$. These co-occurring events were accumulated in time for each spatial location, defining in this way the  joint distribution probability reported in Figure \ref{fig_prob_onset}. Thus this distribution describes how events in the Left Insula trigger events in the Paracentral Right Lobule (positive lags) or vice-versa (negative lags). The distributions in the patient and control groups are compared with randomized cases in which the timing of the onsets in the Insula were randomized (on 500 trials) , preserving the original values of the inter-events times (Figure \ref{fig_prob_onset}). These distributions confirm the decreased influence in both direction found by GC analysis.

\section{Discussion}

The present results add to a large body of evidence indicating  that chronic pain involves dynamical changes that affects normal brain function which in turn can impair cognitive function, including depression, sleeping disturbances and decision-making abnormalities \cite{apka2004,apka2005,baliki2008}. 

The approach used here is derived from two lines of work that together allows for a novel view of the changes in brain effective connectivity. On one side adds strength  to the previous suggestions \cite{tagliazucchi2010a,tagliazucchi2010b,tagliazucchi2012,wu2013} that a few (relatively large) BOLD events can contain substantial information to describe functional connectivity. On the other side, the seed based Granger causality mapping  allows the precise description of the effective connectivity.

The changes in the effective connectivity described in the results section are fully consistent with the current
data \cite{AVAreview}  validating the novel methodology and encouraging a more detailed analysis of other regions of interest described recently \cite{baliki2012} as involved in the transition from acute to chronic pain.

In summary, we show that by deconvolving the hemodynamic response function to spontaneous neural  events it can be demonstrated via Granger Causality that the chronic pain patients exhibits important changes in the Insula's effective connectivity. This novel method can be relevant to understand the overall impact of chronic pain on brain function.

\section*{Acknowledgments}

Work supported by NIH (USA), CONICET (Argentina), BELSPO and UGent BOF (Belgium). ET was
partially funded by the Bundesministerium fur Bildung und Forschung (grant 01 EV
0703) and LOEWE Neuronale Koordination Forschungsschwerpunkt Frankfurt (NeFF).

\section*{References}




\end{document}